\documentclass[runningheads]{llncs}
\usepackage{cite}
\usepackage{amsmath,amssymb,amsfonts}
\usepackage{algorithmic}
\usepackage{graphicx}
\usepackage{textcomp}
\usepackage[]{caption}
\usepackage[labelformat=simple]{subcaption}

\usepackage{listings,color}
\usepackage[dvipsnames,rgb]{xcolor}

\usepackage{float} 
\usepackage[utf8]{inputenc}

\usepackage{makecell}
\usepackage{tablefootnote}
\usepackage{siunitx}
\usepackage{mathtools}
\usepackage{multirow}
\usepackage{enumitem}
\newlist{steps}{enumerate}{1}
\setlist[steps, 1]{label =\roman*:}

\def\BibTeX{{\rm B\kern-.05em{\sc i\kern-.025em b}\kern-.08em
    T\kern-.1667em\lower.7ex\hbox{E}\kern-.125emX}}

\begin{document}

\title{EAGP: An Energy-Aware Gossip Protocol for Wireless Sensor Networks}

\author{Bruno Chianca Ferreira\inst{1} \and
	Vítor Fonte\inst{2} \and
	João Marco C. Silva\inst{3}}
\authorrunning{B. Chianca et al.}
% First names are abbreviated in the running head.
% If there are more than two authors, 'et al.' is used.
%
\institute{
ENAC, ReSCo Group, Telecom, University of Toulouse, Toulouse, France
\email{bruno.chianca@enac.fr} \and
HASLab, INESC TEC, Universidade do Minho, Braga, Portugal\\
\email{vff@di.uminho.pt} \and
HASLab, INESC TEC, Universidade do Minho, Braga, Portugal
\email{joao.marco@inesctec.pt}\\
}

\maketitle

\begin{abstract}
 In Wireless Sensor Networks (WSN), typically composed of nodes with resource constraints, leveraging efficient processes is crucial to enhance the network lifetime and, consequently, the sustainability in ultra-dense and heterogeneous environments, such as smart cities. Particularly, balancing the energy required to transport data efficiently across such dynamic environments poses significant challenges to routing protocol design and operation, being the trade-off of reducing data redundancy while achieving an acceptable delivery rate a fundamental research topic. In this way, this work proposes a new energy-aware epidemic protocol that uses the current state of the network energy to create a dynamic distribution topology by self-adjusting each node forwarding behavior as {\it eager} or {\it lazy} according to the local residual battery. Simulated evaluations demonstrate its efficiency in energy consumption, delivery rate, and reduced computational burden when compared with classical gossip protocols as well as with a directional protocol. 
\end{abstract}

\begin{keywords}
routing, protocol, gossip, iot, energy-aware
\end{keywords}

\section{Introduction}

Wireless Sensor Networks (WSN) are composed of several nodes with a specific purpose of monitoring diverse types of systems and physical phenomena. Typically, these nodes are small and low-cost devices designed to run on limited energy resources. Depending on the deployment scenario, they should not receive maintenance intervention for years, making it imperative to design optimal strategies towards lowering the required power to sense, process, store, and mainly to forward data, as the communication subsystem is predominantly the most demanding \cite{6}\cite{mehmood2017internet}.

The way data is transmitted through a WSN is directly related to the underlying application requirements. When it demands efficient dissemination of the acquired data to all nodes, usually, it is employed some form of epidemic distribution, being the gossip protocol family the most common \cite{boukerche2011routing}. In these protocols, a node immediately forwards messages to all or a subset\footnote{A parameter commonly called {\it fanout}.} of its neighbors, assuring high delivery efficiency and network coverage, though, with the disadvantage of wasting energy with excessive message redundancy. For applications requiring data delivery to a central entity, {\it e.g.}, a base station or a sink, directional routing protocols prevail, as they are efficient in terms of delivery with reduced redundancy. However, they provide lower network coverage, and might lead to battery depletion of nodes in the best path faster, creating energy unbalance, and in extreme cases, completely isolating part of the network.

In this context, this work proposes an Energy-Aware Gossip Protocol (EAGP) able to reduce data redundancy while providing high delivery rate and broad network coverage. The aim is to extend the network lifetime by dynamically optimizing the energy consumed in data dissemination according to the remaining battery level of nodes into the same range. The rationale is to self-adjust which nodes will forward a received message, and the time they wait to do so. In this way, nodes with higher level of residual battery assume an {\it eager} behavior, whilst nodes with lower levels wait and only forward the message in case of failure, {\it i.e.}, {\it lazy} behavior.    

Simulated results evince that, for diverse scenarios, the proposed protocol achieves better performance regarding network longevity and delivery efficiency when compared to classic gossip protocols, including a {\it fanout} version. Moreover, the evaluation also demonstrates promising enhancements compared to a direct routing protocol due to its ability to promptly adapt the distribution topology in case of node mobility or failure.   

An additional contribution from this work is a publicly available framework designed for easy and modular deployment of different protocols, test scenarios, and performance evaluation\footnote{Available at: https://github.com/brunobcfum/pyeagp}. Built over a well-established simulation engine (see Section~\ref{cap:methodology}), this framework might shorten development and optimizations required by specific applications. 

This paper is organized as follows: related work is discussed in Section~\ref{related}; the proposed protocol design goals and rationale are presented in Section~\ref{sec:eagp}; the methodology of tests is presented in Section~\ref{cap:methodology}; the proof-of-concept and the corresponding evaluation results are discussed in Section~\ref{sec:results}; and the conclusions are summarized in Section~\ref{sec:conclusions}.

\section{Related Work}
\label{related}

Over the last decades, several routing protocols for WSN have been designed for a variety of applications and deployment scenarios. An encompassing taxonomy of them is described in \cite{4}, in which, for network structured-based protocols, they are mainly divided into three classes: \textit{flat-based} routing, \textit{hierarchical-based} routing, and \textit{location-based} routing. 

In the \textit{flat-based} routing scheme, every node plays the same role, while in \textit{hierarchical-based} routing, some nodes assume the role of {\it cluster head} and concentrate data before forwarding to another node and eventually to the base station. \textit{Location-based} protocols use the sensors' geographical position to define their role in the routing scheme. Regardless the class, energy consumption has been an active research topic. The predominant approach is to conceive adaptive {\it hierarchical-based} protocols that react to the current state of the network and automatically adjust nodes' role. 

For epidemic protocols, beyond the strategy of forwarding data to only a subset of the nodes' neighbors \cite{5}, some protocols reduce data in the network by resorting to query-based models, {\it e.g.}, SPIN \cite{spin}. In these models, the sink only requests data it is interested in based on metadata announced by the network sensors, which avoid all the nodes continuously transmitting the acquired new data \cite{4}. However, when the payload of the required data is small, the overhead related to advertising and requesting a new measured data affects the protocol efficiency. 

Another strategy is introduced by event-based algorithms, such as Directed Diffusion \cite{dd} and Rumour Routing \cite{3}, in which the propagation of new data ({\it i.e., event}) creates directional routing tables. However, with a negative impact on energy efficiency during the early stages of the network, when no gradient has been created in the directed diffusion and no agent has been issued in the rumor routing. After some time, all nodes will eventually have tables with gradients and routes to events but keeping them increases its computational weight.

Several protocols have introduced energy-aware mechanisms, for instance, hierarchical protocols such as SEP \cite{sep} and TEAR \cite{8368317} resort to a probability model in which nodes with higher energy levels have more chances to be selected as {\it cluster head}. In the latter protocol, in addition to the energy level, it also takes into consideration the network activity. An encompassing analysis of energy-aware strategies on WSN can be found in \cite{jabbar2018}, where it is highlighted the lack of energy-aware solution for {\it flat-based} protocols.

In general, protocols based on shortest path routing, such as MCFA \cite{mcfa}, are efficient in delivering messages to a fixed sink as they ensure a high delivery rate with low message redundancy. Though, as a consequence, they provide limited coverage and resilience since the failure of specific nodes might isolate part of the network. %, which cannot heal itself.  

The above discussion evinces the need for WSN routing protocols able to provide high performance in data delivery and network coverage while reducing the overall required energy. The present work proposes a gossip-based protocol that reduces data redundancy by introducing additional techniques to cancel the forwarding of some messages in a controlled manner, thus enhancing energy efficiency for continuous delivery applications running over {\it flat} topologies.

\section{EAGP: Energy-Aware Gossip Protocol}
\label{sec:eagp}

Aiming to optimize the energy drained by epidemic protocols forwarding duplicate messages, the proposed protocol introduces a new scheme of energy-aware data distribution. It consists of self-adjusting the pace each node forwards message proportionally to its energy level compared to the neighborhood and avoiding retransmission of messages flowing efficiently through the network. Devising such a protocol follows specific design goals, presented in Section~\ref{sec:goals}. The protocol rationale is further detailed in Section~\ref{sec:rationale}.
	
\subsection{Design goals }
\label{sec:goals}

In WSN, designing effective routing protocols requires adjusting the way data is processed and distributed according to the application demands and constraints. For instance, aggregating data in transit reduces overall energy consumption. However, it also might lead to lower efficiency in data delivery, mainly in the presence of node failures or mobility. In this way, devising an energy-efficient gossip protocol able to balance those requirements must follow essential design principles, as described below:  % deixar claro que são objetivos deste protocolo em específico
     
\begin{itemize}
\item the protocol must increase the network lifetime by optimizing the energy consumed in forwarding tasks. Such enhancement is achieved through mechanisms designed to reduce the level of redundant messages traversing the network;
\item it also must ensure a high delivery rate of messages considering both high network coverage and low message losses. A trade-off between the delivery rate and energy-efficiency is acceptable as long as it is an application design choice;
\item another key feature is the balance in energy consumption among all nodes forwarding messages, avoiding the overload of those in the best path, which might lead to network partitioning;
\item considering WSN's constrained devices, the routing algorithm complexity has a direct impact on the energy used by the nodes. Hence, the proposed protocol must comply with the aforementioned goals requiring minimum computational overhead.
\end{itemize}

\subsection{Protocol rationale}
\label{sec:rationale}

In multi-hop topologies, a node typically forwards two types of message, ({\it i}) those containing data acquired locally via sensing events and ({\it ii}) those in which the node is only relaying data between different nodes. For the first group, EAGP follows a continuous delivery flow model, in which the sensors report newly acquired data to a sink periodically \cite{4}. The mechanisms introduced in this work target the latter and consist of self-adjusting the way messages are forwarded according to the battery level of reachable nodes. In this way, nodes with higher residual energy assume an {\it eager} behavior, transmitting messages promptly, which ensures data is distributed with no delay. On the other hand, nodes with lower energy assume a {\it lazy} behavior, meaning they will hold messages longer, and forward them later as a backup in case of failure in the eager path.

The decision between lazy or eager mode is done for each message arriving to a node and takes into consideration the local energy level ({\it i.e.}, $\varepsilon_{lc}$), and the average energy level of all neighbor nodes\footnote{The peer management protocol is beyond the scope of this work.}({\it i.e.}, $\bar{\varepsilon}_{nb}$). The energy level is periodically advertised to all neighbors based on a configurable threshold $\lambda$ ({\it e.g.}, using time or battery variation triggers), preferably through piggybacking strategies.

As presented in Equation~\ref{eq:mode}, if the local residual battery is lower than the neighbors' average, the node assumes the lazy mode for such message, since other nodes with higher battery level will forward it quickly ({\it i.e.}, eager mode). As the energy of eager nodes is consumed, different devices will assume such a role, balancing the overall burden and consequently avoiding the overload of only a subset of nodes. 

\begin{equation}
%\[
\begin{cases}
\varepsilon_{lc} < \bar{\varepsilon}_{nb}  \to \text{{\it Lazy mode}} \\
\varepsilon_{lc} \geq \bar{\varepsilon}_{nb}  \to \text{{\it Eager mode}}
\end{cases}
%\]
\label{eq:mode}
\end{equation}

After defining the forwarding mode for a specific message $M_{id}$, the node follows the underlying operations detailed below. 

\subsubsection{Lazy mode}

\begin{steps}
\item $M_{id}$ is stored in a {\it lazy queue} during a maximum configurable time $\Delta T_{max}$. This value might be tuned according to the application requirements or network dimension; 

\item during this time, if the same message arrives from a different sender, it means the data is flowing across the network. Then, $M_{id}$ is removed from the {\it lazy queue} and discarded;

\item every $\Delta T_{max}$ cycle, the node sends an {\it advertising message} to all neighbors containing the identification of the messages stored in the {\it lazy queue} with $\Delta T_{max}$ exceeded. In this way, nodes that did not receive such a message can request it directly;

\item after a timeout $T_{rec}$, which might also be configured according to the application needs, $M_{id}$ is definitively removed from the {\it lazy queue}.  
\end{steps}

\subsubsection{Eager mode}

\begin{steps}
\item $M_{id}$ is scheduled to be dispatched after the period $\Delta T_{next}$. This time is proportional to the energy level and is defined according to Equation~\ref{eq:tnxt}. \\
\begin{equation}
\Delta T_{next} = \Delta T_{max} - (\Delta T_{max} * \frac{\varepsilon_{lc} - min(\varepsilon_{nb})}{max(\varepsilon_{nb})- min(\varepsilon_{nb})})
\label{eq:tnxt}
\end{equation}

Where,

$\varepsilon_{lc}$ $\rightarrow$ energy level of the node in percentage;   
               
$\varepsilon_{nb}$ $\rightarrow$ list with the energy levels of the node's neighbors in percentage;   

$min(\varepsilon_{nb})$ $\rightarrow$ minimum energy level of the neighborhood; 

$max(\varepsilon_{nb})$ $\rightarrow$ maximum energy level of the neighborhood. 

The rationale behind this equation is to scale the battery level of visible nodes in a value between $0$ and $1$ using the $Min-Max$ normalization technique. Therefore, the eager node with the highest energy will forward $M_{id}$ immediately, while the others will hold it for $\Delta T_{next} \leq \Delta T_{max}$ proportionally to their energy level. This pattern prevents all eager nodes forwarding $M_{id}$ at the same time;   

\item during this time, if the same message arrives from a different sender, the node moves $M_{id}$ to the {\it lazy queue} and follows the underlying process previously described;

\item after $\Delta T_{next}$, $M_{id}$ is forwarded to all neighbors or to a subset of them. 

\end{steps}

\section{Proof-of-concept}
\label{cap:methodology}

Aiming to evaluate the ability to reduce overall energy consumption while providing high performance in message delivery, a comprehensive proof-of-concept assesses the proposed protocol in diverse scenarios. The tests consist of comparing its results with other well-established protocols in a simulated environment, which ensures that no external variables can tamper with the analysis. Details regarding the underlying methodology are presented along this section. 

\subsection{Compared protocols}
\label{simulated_protocols}

A reference implementation of EAGP (detailed in Section~\ref{sec:rationale}) is compared with two widely deployed versions of the Gossip protocol and a directional protocol, as described below:

\begin{itemize}

    \item Gossip - gossip protocols are usually implemented with some variety of fanout as an optimization to reduce message redundancy. The strategy relies on forwarding messages only to a subset of node's neighbors instead of all of them, as in the standard gossip. In this work, EAGP is compared with both the standard and fanout versions ({\it i.e.,} Gossip FO). Based on the simulation scenarios (described in Section~\ref{sec:test:scn}), the fanout size is experimentally set to $3$.
    
    \item MCFA - it is a directional protocol, meaning that messages transmitted from the creator node to the sink node traverse the network through a unique best path, which prevents message redundancy \cite{mcfa}. Since it can be implemented with the continuous delivery traffic model, it might provide valuable comparative insights regarding EAGP's performance.  
    
\end{itemize}

Following the parameterization strategy adopted in Gossip FO, the EAGP queue timeouts are experimentally defined according to the simulation scenarios as $\Delta T_{max} = 10 sec$ and $\Delta T_{rec} = 2*\Delta T_{max}$. Other configurable parameters are the message {\it TTL}, which is set as two times the topology diameter for all protocols (see Section~\ref{sec:test:scn}), and $\lambda$, set as 10\% of energy variation.

\subsection{Comparative parameters}

As described in Section~\ref{sec:goals}, EAGP's main goal is to balance the network lifetime with acceptable performance in data delivery. Thus, assessing the underlying achievements involves a comparative statistical analysis based on the following metrics. % \cite{Yitayal2015}: 

\begin{itemize}

    \item Network longevity: a WSN that implements an energy-efficient protocol to distribute data is expected to live longer in the sense that it will take longer until the last node is able to deliver data to a sink. In this way, the energy used by the network during the simulation time is recorded and compared based on the energy model described in Section~\ref{sec:model};

    \item Delivery rate: EAGP's performance is assessed by measuring whether messages created by a node are delivered to all active nodes in the network. Moreover, during the simulations, one node will always be selected as a sink and taken as a reference. This strategy allows the protocol evaluation for different types of applications.

    \item Data redundancy: using a gossip-based distribution process creates message replicas that follow different paths towards the sink. Hence, the same data might be received by the sink more than once. Redundancy can be valuable to ensure high reliability, but it also can lead to energy waste. Comparing the delivery rate with the number of duplicated messages arriving at the sink provides a valuable indication of EAGP's global performance;
    
    \item Energy efficiency - after each simulation, it is estimated the average of energy consumed per message delivered to the sink. The least energy required, the more energy-efficient the routing algorithm is.
    
\end{itemize}

\subsection{Test Scenarios}
\label{sec:test:scn}

\begin{figure*}[!h]
    \centering
    \begin{subfigure}[b]{0.4\textwidth}
        %\centering
        \includegraphics[width=0.8\linewidth]{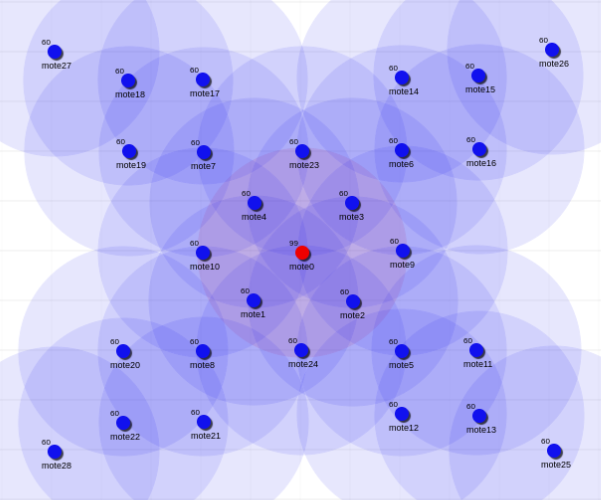}
        \caption{Symmetrical.}
        \label{subfig:symmetrical}
    \end{subfigure}
    \qquad
    \begin{subfigure}[b]{0.4\textwidth}
        %\centering
        \includegraphics[width=1.1\linewidth]{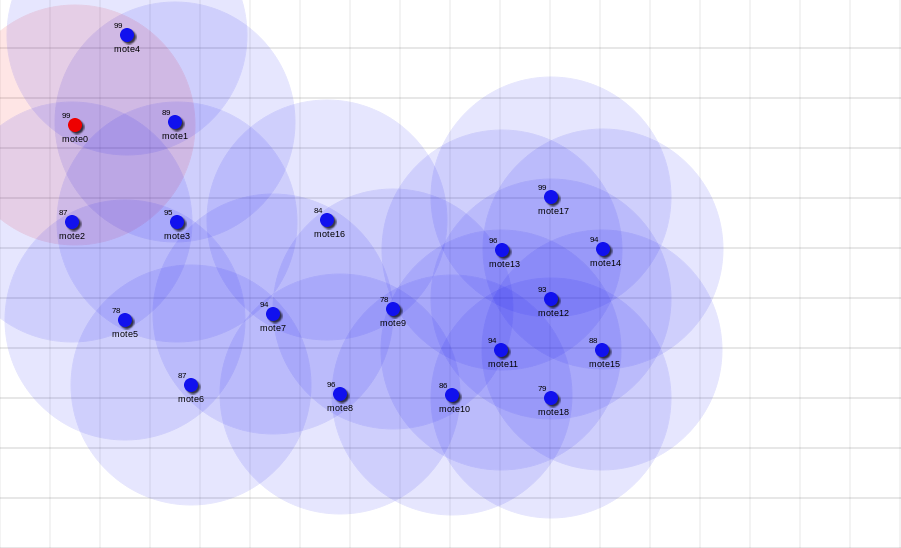}
        \caption{Asymmetrical.}
        \label{subfig:asymmetrical}
    \end{subfigure}
    \qquad
     \begin{subfigure}[b]{0.4\textwidth}
        %\centering
        \includegraphics[width=1\linewidth]{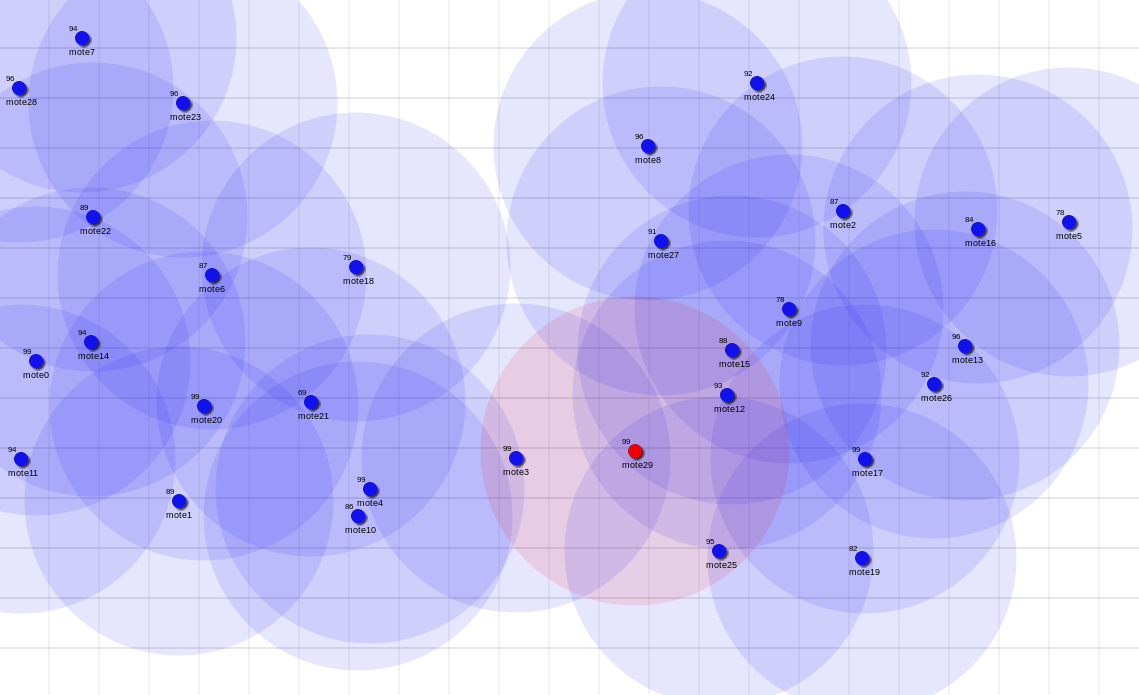}
        \caption{Random.}
        \label{subfig:random}
    \end{subfigure}
    \caption{Network topologies.}
    \label{fig:topologies}
\end{figure*}

Considering the diversity of environments in which WSN is typically deployed, it is essential for its underlying protocols to be tested under different topologies and evaluation approaches.  Therefore, EAGP is comparatively assessed for a generic continuous delivery application operating in three topologies:
 
 \begin{itemize}
 
\item Symmetrical: as presented in Figure~\ref{subfig:symmetrical}, this topology has the sink node in the middle, and all the other nodes equally distributed around it with a maximum radius of 5 hops. It represents the best-case scenario in terms of efficiency, as all nodes are symmetrically distributed with no overlap in signal coverage;

\item Asymmetrical: it has the sink node in the extreme left and all the other nodes distributed to the right, with the maximum distance of 9 hops (see Figure~\ref{subfig:asymmetrical}). In this topology, some nodes are mandatory paths in the way to the sink, for instance, those closest to it, which implies  their energy depletion isolates the whole network;

\item Random: as presented in Figure~\ref{subfig:random}, it has the sink node in the middle, and all the other nodes randomly distributed around it, with the maximum distance of 11 hops. Note that \textit{mote3} connects the whole cluster on the left side to the sink, so this node is essential to this cluster. This topology is closer to a real case since it is non-deterministic and includes overlapping signal coverage. 
\end{itemize} 
 
The analysis takes into consideration three simulation scenarios, {\it i.e.}, {\it steady-state}, {\it end of life}, and {\it mobility}. The steady-state represents a snapshot of the protocol behavior during the majority of the network's lifetime, when all nodes are active and none has its battery depleted before the simulation be concluded. Thus, the test scenario is set in a way that all nodes have an initial large amount of remaining energy available\footnote{For all scenarios, each node starts with different remaining battery to simulate a heterogeneous state.}.

Contrarily, in the end of life scenario, nodes are configured to start with a smaller amount of energy. It aims at evaluating the behavior when the network evolves from a steady-state to the point in which some nodes start do fail due to battery exhaustion. The mobility scenario shares the initial setup with the steady-state scenario. However, it is added mobility to the nodes in order to observe the protocol's versatility in adapting to topology changes during the network operation. Since some algorithms have a start-up phase, the movement is only introduced after this stage and consists of a random walk model updated every second \cite{8741010}.

It is important to highlight that all protocols are assessed under the exact same scenarios, including the initial amount of energy in each node, the energy consumption model (presented in Section~\ref{sec:model}), and the mobility pattern.

\subsection{Simulation model}
\label{sec:model}

The comparative analysis is performed following the well-established ESP8266 energy model~\cite{esp:8266}. Thus, all the considered protocols are implemented in the CORE simulator~\cite{ahrenholz2010comparison}, resorting to a modular and publicly available framework developed into this work scope.  

In this simulation environment, the sensors' schedulers are based on the host time, meaning that time-related processes and test measurements share a global clock.  Table~\ref{tab:parameters} details the underlying energy model along with fixed parameters used across the simulations. The results discussed in Section~\ref{sec:results} correspond to the average of five simulations for each scenario.

\begin{table}[!h]
    \centering
    %\captionsetup{justification=centering, labelsep=newline}
    \caption{Energy model and fixed parameters.}
    \label{tab:parameters}
    \begin{tabular}{|l|l||l|l|}
    \hline
    Deep sleep  & \num{1e-5}$A$   & Battery voltage     & 3.7$V$                            \\ \hline
    Modem sleep & \num{1.5e-3}$A$ & $TX$ time           & \num{30e-3}$sec$ \\ \hline
    Awake       & \num{8.1e-3}$A$ & $RX$ time           & \num{40e-3}$sec$ \\ \hline
    $TX$ current        & \num{1.7e-2}$A$ & Badwidth            & \num{54e6}$bps$  \\ \hline
    $RX$ current        & \num{5.6e-3}$A$ & One-hop delay & \num{5e-3}$sec$  \\ \hline
    Sensor energy       & \num{1.1e-9}$J$ & Jitter / Error      & 0                                 \\ \hline
    \end{tabular}
\end{table}

Another important aspect of the simulated scenarios is the continuous delivery model adopted. In this sense, for the analyzed protocols, all nodes in the network produce new data in a frequency randomly bounded between 15 and 50 seconds.  
 
\section{Results and Discussion}
\label{sec:results}

Considering the balance between energy consumption and data delivery the main goal in EAGP's design, Table~\ref{tab:summary} presents the summary of simulation results for the steady-state scenario. It reveals the protocols' behavior in most of their operation time. In such scenario, a direct comparison between gossip optimizations, {\it i.e.}, EAGP and Gossip FO, shows a significant improvement in energy cost per packet delivered to the sink node by the first. This is possible due to the reduction of replicated data traversing the network without affecting the overall delivery rate.

\begin{table}[H]
   % \captionsetup{justification=centering, labelsep=newline}
	\centering
	\caption{Overall results for the steady-state scenario.}
	\label{tab:summary}
\begin{tabular}{l|l|l|l|l|}
\cline{2-5}
                                           & Gossip    & Gossip FO & EAGP   & MCFA  \\ \hline
\multicolumn{5}{|c|}{Symmetrical topology}                                            \\ \hline
\multicolumn{1}{|l|}{Energy (J)}           & 23,854    & 8,978     & 8,904  & 2,422 \\ \hline
\multicolumn{1}{|l|}{Delivery rate (\%)}   & 99.73     & 84.96     & 91.00  & 99.73 \\ \hline
\multicolumn{1}{|l|}{Efficiency (J/pkt)}   & 2.43      & 1.07      & 0.99   & 0.25  \\ \hline
\multicolumn{1}{|l|}{Data redundancy \tablefootnote{Data redundancy is the average number of repeated packets received.}} & 8.26      & 3.04      & 2.71   & 1.88  \\ \hline
\multicolumn{5}{|c|}{Asymmetrical topology}                                           \\ \hline
\multicolumn{1}{|l|}{Energy (J)}           & 771,238   & 101,212   & 9,297  & 5,216 \\ \hline
\multicolumn{1}{|l|}{Delivery rate (\%)}   & 86.82     & 70.61     & 75.30  & 99.71 \\ \hline
\multicolumn{1}{|l|}{Efficiency (J/pkt)}   & 161.65    & 23.66     & 1.98   & 0.87  \\ \hline
\multicolumn{1}{|l|}{Data redundancy } & 139.82    & 11.68     & 1.76   & 4.35  \\ \hline
\multicolumn{5}{|c|}{Random topology}                                               \\ \hline
\multicolumn{1}{|l|}{Energy (J)}           & 1,437,968 & 186,740   & 12,980 & 3,971 \\ \hline
\multicolumn{1}{|l|}{Delivery rate (\%)}   & 70.66     & 96.88     & 90.55  & 99.72 \\ \hline
\multicolumn{1}{|l|}{Efficiency (J/pkt)}   & 224.19    & 18.51     & 1.37   & 0.39  \\ \hline
\multicolumn{1}{|l|}{Data redundancy } & 308.63    & 34.27     & 2.48   & 2.21  \\ \hline
\end{tabular}
\end{table}

\begin{figure}[H]
	\centering
	\includegraphics[width=0.7\textwidth]{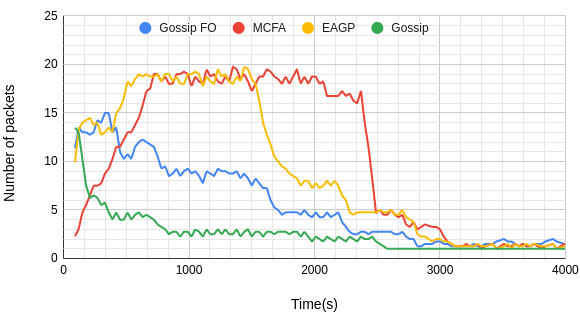}
	\caption{Network longevity.}
	\label{fig:eol_asym}
\end{figure}

The major impact of such optimization is a larger network lifetime. As presented in Figure~\ref{fig:eol_asym} for the {\it end of life} scenario in the asymmetrical topology, EAGP can provide a higher number of packets delivered to the sink node, and during a longer period when compared with both Gossip protocols.

\begin{figure}[h]
	\centering
     \includegraphics[width=0.7\textwidth]{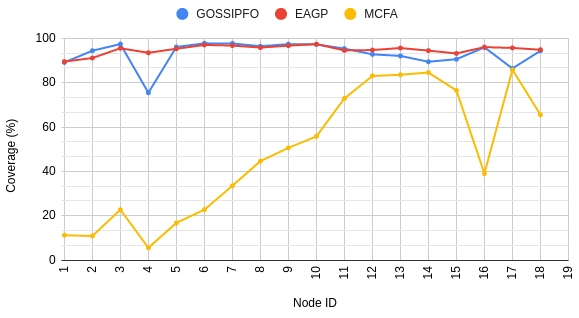}
	\caption{Network coverage in asymmetrical topology.}
	\label{fig:coverage}
\end{figure}

As expected for a directional protocol, MCFA outperforms all epidemic variations in terms of energy consumption for static scenarios. However, it is interesting to note how close the results achieved by EAGP are, mainly when considering its higher network coverage (see Fig.~\ref{fig:coverage}). Contrarily, a pure gossip protocol demands significantly more energy in consequence of the number of replicated messages, particularly in asymmetrical and random topologies. Though, this higher replication does not always represent a better performance in data delivery, as observed in the random topology (see Table~\ref{tab:summary}). 

\begin{figure}[h]
	\centering
     \includegraphics[width=0.7\textwidth]{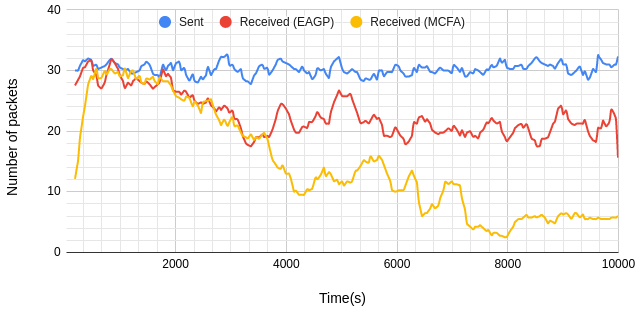}
	\caption{Protocol performance in mobility scenario.}
	\label{fig:eol_geral_mob}
\end{figure}

When considering a scenario with moving nodes, MCFA faces a considerable performance drop, and EAGP surpasses all compared protocols. Figure~\ref{fig:eol_geral_mob} demonstrates it by presenting the relation between the number of messages sent from all nodes and the total of unique messages arriving at the reference node. In this case, MCFA's delivery rate is 46\%, while EAGP reaches 68\%. The standard Gossip and the fanout version ratio are 56\% and 65\%, respectively.

For applications requiring high network coverage, EAGP is able to reach significantly better performance, as shown in Figure~\ref{fig:coverage}. This analysis ratifies its effectiveness in covering a broader number of nodes with a lower cost per message. Comparatively, in MCFA, data generated in some nodes reach less than 20\% of the network, while EAGP reaches, on average, more than 90\% of the network. This performance is similar to Gossip FO but consuming 87\% less energy and with 84\% less redundancy.   

\begin{figure}[H]
	\begin{subfigure}[b]{0.5\textwidth}
		\centering
		\includegraphics[width=1\linewidth]{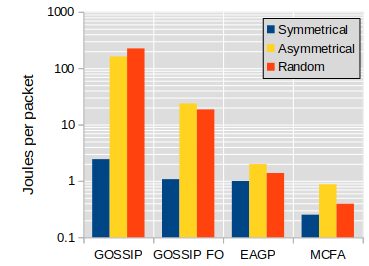}
		\caption{Energy per message}
		\label{fig:energy_eff}
	\end{subfigure}
	\hfill
	\begin{subfigure}[b]{0.5\textwidth}
		\centering
		\includegraphics[width=1\linewidth]{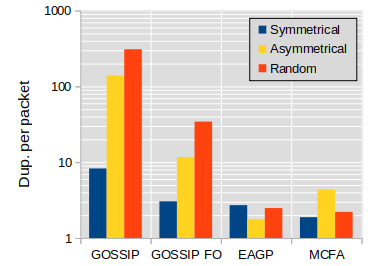}
		\caption{Message redundancy}
		\label{fig:delivery_eff}
	\end{subfigure}
	\caption{Protocol efficiency in different scenarios.}
	\label{fig:fwd_disc}
\end{figure}

Moreover, the ability to self-adjust data distribution according to the network dynamics leads EAGP to the smallest performance variation across all tested scenarios, highlighting its suitability to heterogeneous environments. This can be observed in Figure~\ref{fig:fwd_disc}, which also evinces the relation between the volume of redundant messages traversing the network and the total amount of energy consumed.        

\begin{figure}[H]
	\centering
	\begin{subfigure}[b]{0.45\textwidth}
		%\centering
        \includegraphics[width=1\linewidth]{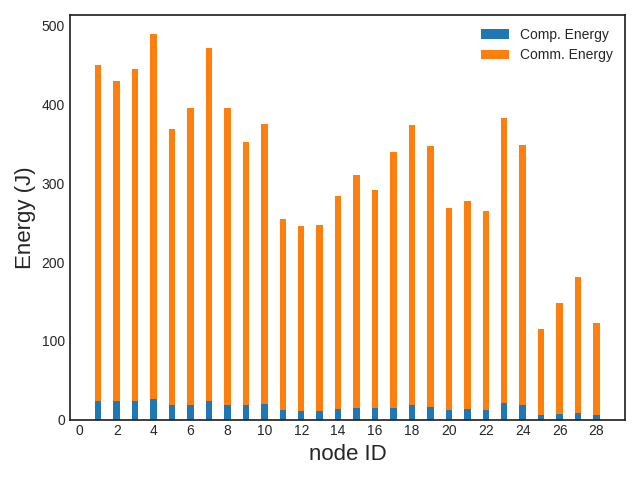}
		\caption{Gossip FO}
		\label{fig:energy_prof_gfo}
	\end{subfigure}
	\hfill
	\begin{subfigure}[b]{0.45\textwidth}
		%\centering
        \includegraphics[width=1\linewidth]{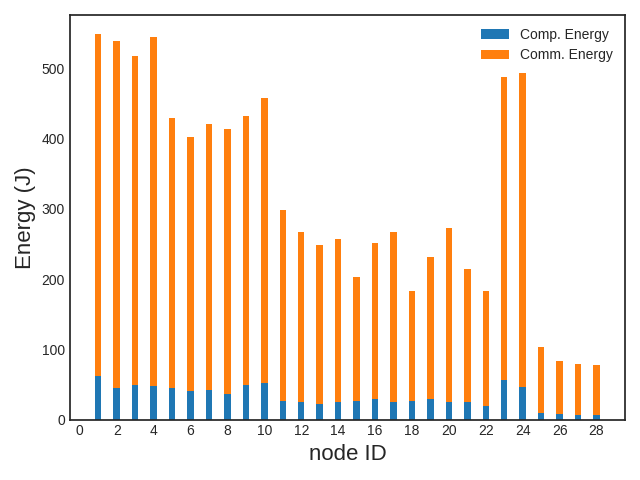}
		\caption{EAGP}
		\label{fig:energy_prof_eagp}
	\end{subfigure}
	\caption{Energy consumption profile.}
	\label{fig:energy}
\end{figure}

Designing a WSN routing protocol based on adaptive processes may demand additional energy in computation not related to communication or sensing tasks. For instance, while the fanout version of the gossip protocol chooses random neighbors to forward messages, in EAGP, there is an additional computational footprint of updating node's state, calculating $\Delta T\textsubscript{NEXT}$, and maintaining different queues for message scheduling. Figure~\ref{fig:energy} details the energy consumption profile of each node for the same scenario. In this sense, the overhead required by EAGP is compensated through the reduction of redundant data transmitted along the network's lifetime.  

\section{Conclusions}
\label{sec:conclusions}

This work has introduced a new gossip-based protocol able to self-adjust data forwarding according to the residual energy of reachable nodes. The strategy consists of reducing the amount of redundant data traversing the network by canceling the forwarding task of messages flowing the network efficiently for nodes with lower energy.

Resorting to a simulation-based prototype, a proof-of-concept has demonstrated promising results when comparing the proposed protocol with well-established gossip protocols regarding network longevity and data delivery performance. The achieved results are even comparable with a non-epidemic protocol, which ratify its overall efficiency. As future work, the tests will consider deploying sensors based on ESP8266 for a specific continuous delivery application, namely, a temperature and humidity monitor.

%\section*{Acknowledgments}
%\small{This work is financed by National Funds through the Portuguese funding agency, FCT - Fundação para a Ciência e a Tecnologia within project UIDB/50014/2020.}

\bibliography{IEEEabrv,main}

\end{document}